
\input phyzzx.tex

\hoffset=0.2truein
\voffset=0.1truein
\hsize=6truein
\def\TITLEPAGE{\frontpagetrue}
\def\CALT#1{\hbox to\hsize{\tenpoint \baselineskip=12pt
        \hfil\vtop{
        \hbox{\strut CALT-68-#1}}}}

\def\CALTECH{
        \address{California Institute of Technology,
Pasadena, CA 91125}}

\def\AUTHOR#1{\vskip .2in \centerline{#1}}

\def\ABSTRACT#1{\vskip .2in \vfil \centerline{\twelvepoint
\bf Abstract}
        #1 \vfil}
\def\ENDTITLEPAGE{\vfil\eject\pageno=1}

\REF\complementarity{L. Susskind, L. Thorlacius, and J. Uglum, Phys.
Rev. {\bf D48} (1993), 3743; L. Susskind and L. Thorlacius, Phys. Rev.
{\bf D49} (1994), 966.}
\REF\wormhole{V. Frolov and I. Novikov, Phys. Rev. {\bf D48}
(1993), 1607.}
\REF\remnants{Y. Aharanov, A. Casher, and S. Nussinov, Phys. Lett.
{\bf B191} (1987) 51.}

\TITLEPAGE
\CALT{1931}
\bigskip            
\titlestyle {Traversable Wormholes and Black Hole Complementarity}
\AUTHOR{Daniel Gottesman}
\CALTECH
\ABSTRACT{Black hole complementarity is incompatible with the
existence of traversable wormholes.  In fact, traversable wormholes
cause problems for any theory where information comes out in the
Hawking radiation.}

\ENDTITLEPAGE

\eject

One school of thought on the black hole information-loss problem
says that information encoded on infalling matter comes out subtly
encoded in the Hawking radiation, which is only approximately
thermal.  The main problem faced by this proposal is that it produces
a duplication of information: for a large black hole, the curvature at
the horizon is small, so Alice should be able to cross the horizon
unharmed.  In addition to the original Alice, there will be a copy,
who I will call Abby, encoded in the Hawking radiation.  This sort of
duplication of information causes evolution of pure states into mixed
states, which is precisely the problem that was to be avoided.

Black hole complementarity is a recent proposal by Susskind {\it et
al.}$^{\complementarity }$ to avoid this difficulty.  It states that the
duplication of information is visible only to an unphysical
superobserver.  Attempts to bring the two copies together will either
fail completely due to the causal structure of the black hole or will
require Planck scale energies.

However, Frolov and Novikov$^{\wormhole }$ showed that
traversable wormholes could be used to significantly change the
causal structure of a black hole with only small perturbations to the
Schwarzschild metric.  By letting one mouth cross the apparent event
horizon while the other remains outside, we create timelike paths
that cross the apparent horizon, pass through the wormhole, and
then escape to infinity.  If such a path does not pass very close to the
singularity, at no time are large energies involved.

This causal structure allows the violation of black hole
complementarity, and in fact causes problems for any scheme that
has infalling information duplicated in the Hawking radiation.
Suppose Alice follows one of these time-like paths.  Bob, who has
waited outside the black hole, can then meet Alice as she leaves the
wormhole mouth and wait with her until the black hole has
completely evaporated.  By studying the Hawking radiation, Bob and
Alice can reconstruct Abby and allow her to meet her clone, in
violation of complementarity.

There are only two basic ways to avoid this conclusion while still
permitting traversable wormholes.  Either the wormhole pinches off,
preventing Alice from escaping from the black hole; or the
introduction or use of the wormhole distorts the Hawking radiation,
preventing Abby from appearing.  In the case where the black hole
mass and radius are much larger than the mass and radius of a
wormhole mouth, neither of these is possible.

In order to prevent Alice from escaping back across the event
horizon, the wormhole must pinch off immediately upon crossing the
horizon.  If the wormhole is ever traversable with a mouth on each
side of the horizon, Alice can escape the black hole and confront
Abby.  Classical effects will not cause it to pinch off, but perhaps
some quantum mechanical effect enters that closes the wormhole.
However, the curvature due to an arbitrarily large black hole is
arbitrarily small at the horizon.  If one mouth of the wormhole is just
inside the horizon and the other is just outside, a region containing
both wormhole mouths is only infinitesimally different from flat
Minkowski space with a wormhole.  If the wormhole can exist at all,
it should not pinch off at the black hole's horizon.  It may pinch off at
a later time, but it will already be too late to preserve
complementarity.

It is similarly impossible for the wormhole to distort the Hawking
radiation sufficiently to completely destroy Abby.  Although the
causal structure of black hole plus wormhole will be significantly
different than of just a black hole, the metric will be very similar in
the case of a small wormhole mouth.  The Hawking radiation should
therefore change very little when a wormhole mouth is dropped into
it.  It cannot, for example, be generated just outside the true horizon,
which in the presence of the wormhole will be drastically different
than the apparent horizon,\foot{The true horizon is generated by null
rays which cannot escape, even through the wormhole, while the
apparent horizon is the same as it would be without the
wormhole.$^{\wormhole }$} for then it could only escape through the
wormhole; this would result in a drastic change in the Hawking
radiation as seen at infinity.  Also, sending an object through the
wormhole cannot affect the Hawking radiation at all, since the outer
surface of the apparent horizon, where the Hawking radiation is
generated, is spatially separated from the mouth inside the horizon.

The radiation can therefore only be perturbed slightly, but in order
to preserve black hole complementarity, a great deal of information
needs to be erased.  While Alice herself need not contain much
information, suppose Jack and Jill cross the event horizon nearby.
We can assume Alice, Jack, and Jill are in independent, uncorrelated
states, so the amount of information available from all three is
exactly the sum of the information available from the three as
individuals.  The wormhole has no way of telling how many or which
of the three will pass through it, so it needs to destroy not only
Abby, but also Jack and Jill's copies.  Instead of only three people, we
could have chosen an arbitrarily large number, so long as they can fit
in the past light cone of the wormhole mouth.  However, the
wormhole is only slightly perturbing the Hawking radiation, so it
should not be able to erase an arbitrarily large amount of
information.

Even if the wormhole is somehow able to erase all that information, a
new problem arises, or rather an old one reappears.  Alice crosses
the horizon, leaving no Abby behind due to the influence of the
wormhole.  Suppose, though, that Alice changes her mind and decides
not to enter the wormhole.  Neither Abby nor Alice escapes the
evaporation of the black hole, and Alice's information is lost to poor
forlorn Bob.

In fact, a similar problem afflicts any theory where information
comes out in the Hawking radiation, even without complementarity.
Suppose Bob, no doubt wondering what happened to Alice, decides to
enter the black hole through the wormhole to look for her.  Again,
Bob's passage cannot change the Hawking radiation if neither mouth
is at the horizon.  Bob never crossed the horizon, so no copy of him
can be made.  If Bob does not manage to leave back through the
wormhole, but instead waits to hit the singularity, his information is
lost completely to Cheryl, who remained outside waiting for them
both.

Bob can have mass much greater than the Planck mass, so we will
eventually be left with a small black hole containing a large amount
of information.  This is not allowed in theories where information
comes out in the radiation, but is not a problem for
remnants$^{\remnants }$ or for theories where the information is
stored near the singularity and comes out at the end.

The only remaining resolution that permits black hole
complementarity is the non-existence of traversable wormholes.  It
is not sufficient for it to be impossible to create a wormhole where
there was none before -- we can consider a universe that started out
with traversable wormholes.  This choice of initial condition should
not affect the resolution of the information-loss problem.  Therefore,
no theory can simultaneously permit traversable wormholes and
black hole complementarity.

I would like to thank John Preskill for many helpful conversations.
This material is based upon work supported under a National Science
Foundation Graduate Research Fellowship and by the U.S. Dept. of
Energy under Grant No. DE-FG03-92-ER40701.

\refout
\bye